# Insights into regularity of 2D 3$d$ transition metal monocarbides formation

*K.V. Larionov*[1,2*], *G. Seifert*,[3] *P.B. Sorokin*[1,2]

[1] National University of Science and Technology MISiS, 4 Leninskiy prospekt, Moscow, 119049, Russian Federation

[2] Moscow Institute of Physics and Technology, Institute lane 9, Dolgoprudniy, Moscow region, Russian Federation

[3] Technische Universitaet Dresden, Bergstr. 66b, Dresden, Germany

* konstantin.larionov@phystech.edu

**ABSTRACT.** Recently several theoretical predictions were made about 2D planar FeC, CoC, NiC, and CuC while their bulk phases still remain unknown. Here, we present generalization of 2D family of 3$d$ transition metal monocarbides (TMC) by searching their stable configurations with DFT methods and evolutionary algorithm. It is found that in the TMC row (TM = Sc-Cu) a tendency of 3D rocksalt phase formation is monotonously interchanging by 2D phase appearance, namely planar orthorhombic TMC characterized by carbon dimers inside metal hexagons. Among them, orthorhombic CoC and FeC monocarbides would be likely formed rather than any other 2D metal carbide phase or metal/graphene interface.

**INTRODUCTION.** A wide range of bulk structures is highly desired to be synthesized as an ultrathin film or even atomically thin monolayer that may bring to light novel properties as was previously made for the 2D silicon (silicene), boron (borophene) and other two-dimensional phases. [1] Indeed, recent high-throughput investigations devoted to the discovery and prediction of the prospective 2D materials [2,3] unambiguously indicate how scarce is our knowledge and how many compounds are located at the frontier of current researches. Among them, pure metal films and their compounds stimulate their research due to promising electronic and magnetic properties that are applicable for nanoelectronics, spintronics and catalytic purposes. [4–9]

Intercalation of a lithium into bilayered graphene uncovered ultrathin and superdense 2D Li phase.[10] Monoatomic Au [11] and Mo-based [12,13] films were fabricated under electronic beam. As for 3$d$ transition metal (TM) based films, in one of the pioneering works, the formation of 2D square Fe cluster inside graphene pore was reported. [14] This result opened up a discussion about possible formation of iron-based 2D materials such as planar FeC [15] and FeO. [16] A similar square-like net was also discovered for 2D CuO suspended on a graphene substrate both experimentally and from DFT calculations, [17] while pure Cu cluster was found to form a hexagonal lattice. In Ref. [18], a tetragonal corrugated lattice of iron monocarbide ($t$-FeC) was predicted, similar to the previously known 2D phases of TiC [19] and YN, [20] as well as potentially stable planar orthorhombic phase ($o$-FeC) was proposed. Noteworthy, this $o$-FeC lattice is characterized by 5- and 7-coordinated carbon and iron atoms, respectively, that can be represented as a net of carbon dimers inside metal hexagons. The same orthorhombic phase was also predicted for CoC, [21] NiC and CuC. [22]

Such extension of 2D phases of 3$d$ transition metal monocarbides (TMC) is particularly noticeable since no comprehensive data is available for their bulk phases. Experimental studies of the first half of 3$d$ metal series are scarce, but there are reports about the formation of NaCl-type (ScC, [23] TiC, [24] VC, [24] CrC [25,26]) or ZnS-type (MnC [27]) bulk crystals. Numerous theoretical papers propose the formation of the rocksalt phase for the rest four TMC as well (TM = Fe, Co, Ni, Cu [28,29]), while due to another study only FeC may exist under a pressure of about 54 GPa. [30] However, since the synthesis of these four monocarbides have not yet been fairly reported so far, a question about their possible formation in bulk phase and energy favorable structure is still open as discussed below in detail.

Lack of experimental reports on bulk 3$d$ TMC formation together with recent theoretical predictions of promising 2D TMC phases raises a suggestion that some metal monocarbides might form specific planar structures rather than 3D crystals. In particular, for the reported 2D orthorhombic phases of FeC, CoC, NiC, and CuC, [18,21,22] further comprehensive studies of their stability is necessary, including an analysis of binary compositional phase TM-C diagrams.

Here, we present a systematic investigation on the stability of two-dimensional 3$d$ transition metal monocarbides amid experimentally known bulk phases and recent theoretical predictions. In the first part of the paper, different 2D TMC phases are compared. In particular, it was shown that $t$-TMC phase (TM = Sc, Ti, V, Cr, Mn, Fe) is metastable and may exist only as isolated monolayer, while $o$-FeC, $o$-CoC, $o$-NiC, and $o$-CuC are energy favorable in comparison with the earlier theoretical propositions of rocksalt phase. A further study demonstrates that such orthorhombic monolayers can be stacked into an energetically favorable bulk crystal. An analysis of the binary compositional phase diagrams shows that only *t*-FeC, *o*-FeC, *o*-CoC monocarbides as well as $Co_2C$ and $Ni_2C$ monolayers might be thermodynamically stable, while copper carbides are intrinsically unstable against decomposition.

**METHODS.** All calculations were performed within the generalized gradient approximation in the Perdew–Burke–Ernzerhof parameterization. We used the projector augmented wave method [31] approximation with periodic boundary conditions implemented in Vienna Ab-initio Simulation Package. [32–35] Plane-wave energy cut-off was set to 400 eV. To calculate the equilibrium atomic structure, the Brillouin zone was sampled according to the Monkhorst–Pack scheme [36] with a 10×10×10 and 12×12×1 grid in the k-space for bulk and films (monolayers) calculations, respectively. A structural relaxation was performed until forces acting on each atom became less than $10^{-4}$ eV/Å. DFT-d3 correction method [37] was taken into account when describing interlayer interactions. To avoid spurious interaction between the neighboring images while calculating monolayers and slabs, a translation vector along non-periodic direction was set to be greater than 15 Å. Searching of stable compounds was performed with evolutionary algorithm as implemented in USPEX code. [38,39] In order to distinguish fully planar and possibly corrugated 2D structures, thickness constraints of 0.8 Å and 1.5 Å were used, respectively.

**RESULTS AND DISCUSSIONS.**

To get a broader view on the 2D metal carbide formation, we have investigated two-dimensional monocarbides of nine 3*d* transition metals (Sc, Ti, V, Cr, Mn, Fe, Co, Ni, Cu). Four different lattice structures were taken into account. Planar orthorhombic (*o*-TMC: e.g. *o*-FeC, [18] *o*-CoC, [21] and *o*-CuC [22]) and corrugated tetragonal lattices (*t*-TMC: e.g. *t*-FeC [18]) are known by now from *ab initio* predictions. In addition, for a wider comparison, we have chosen hexagonal graphene-like (*g*-TMC: e.g. *g*-ZnO [40]) and planar square (*s*-TMC: 2D CuO-like [17]) lattices. Each of the four lattices were considered in several magnetic configurations: non-magnetic (NM), ferromagnetic (FM) and few antiferromagnetic ones (AFM1, AFM2, etc. depending on the unit cell symmetry, see Figure S1). To avoid any artifact of the unit cell size on the results, all structures were taken as $TM_4C_4$.

The calculated complete data set is represented in Table S1, where empty cells mean that the ground state of the corresponding structure has not been found during DFT optimization or one of the AFM configurations is identical to another one. From Table S1 one can conclude that no evidence for the formation of hexagonal *g*-TMC or square *s*-TMC structures was found. Therefore, these two lattices were excluded from further discussion.

To estimate the probability of 2D TMC formation, we have calculated the energy of the corresponding rocksalt cubic structure as reported experimentally and theoretically (see introduction), and then compared it with the lowest energy we obtained for the 2D structures (i.e. either *t*-TMC or *o*-TMC). The energy difference between two-dimensional phase and bulk structure for the 3*d* TMC row is shown in Figure 1.

One can see that *t*-TMC structures (blue region) are less favorable than the cubic (*rs*) phase, but when it is interchanged by *o*-TMC, the energy difference becomes negative, which means that 2D TMC phase is more energy favorable than corresponding rocksalt structure. In addition, it should be noted that the energy difference for FeC is close to zero. This correlates well with the fact that *t*-FeC and *o*-FeC have similar energies, [18] and both structures are considered further. For details, two energy differences (i.e. [(*t*-TMC) - (*rs*-TMC)] and [(*o*-TMC) - (*rs*-TMC)]) are shown separately in Figure S2.

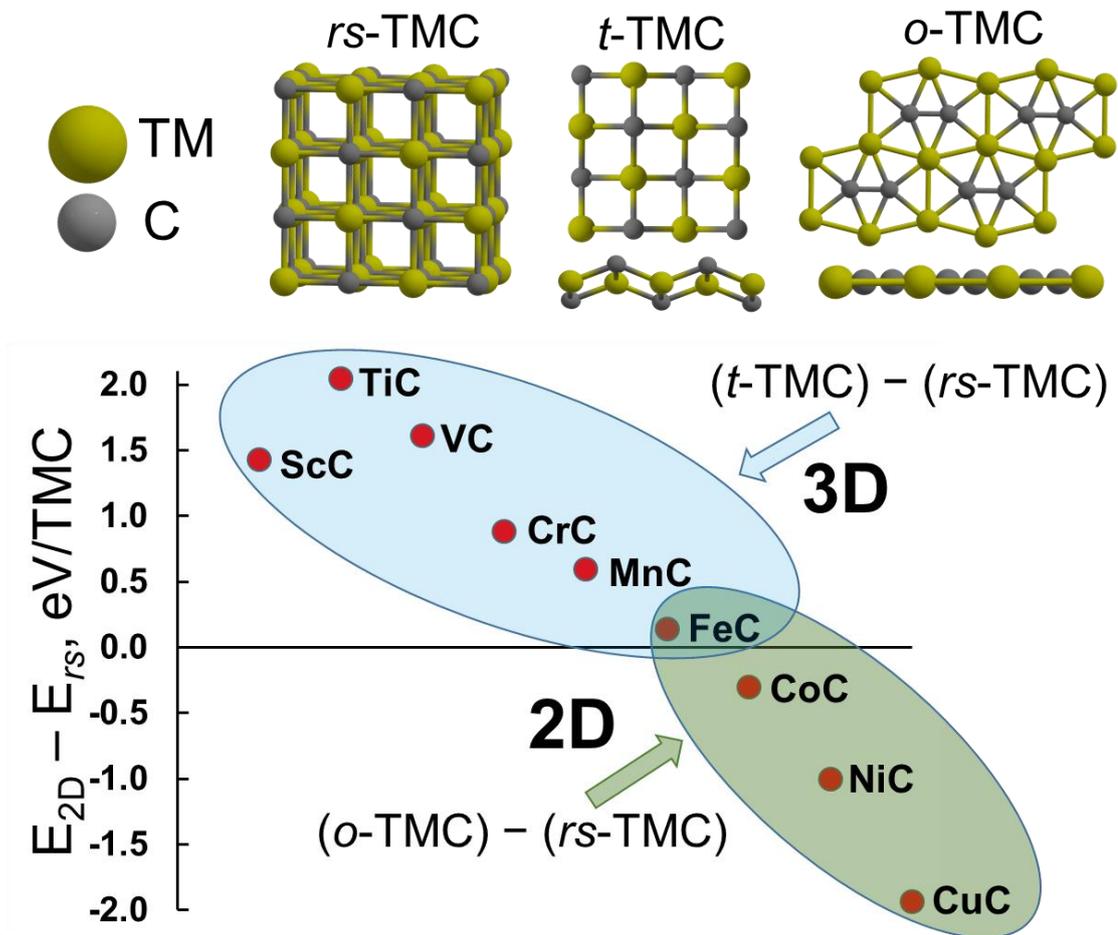

Figure 1. Calculated energy difference between the energetically most favorable 2D TMC phase (either *t*-TMC or *o*-TMC, indicated by blue and green ellipses, respectively) and *rs* bulk phase. The structures of bulk *rs*-TMC, 2D *t*-TMC and *o*-TMC phases are shown above, where olive and grey spheres correspond to transition metal and carbon atoms, respectively.

The origin of the decrease of rocksalt phase stability is thoroughly discussed in the literature. [30,41,42] Generally, rocksalt phase of the transition metal carbides is denoted as interstitials of carbon atoms into TM-sublattices. Two main parameters influence the stability of such compounds: the metal–carbon bond strength which is reducing from the left to the right in 3*d* transition metal row, and the ratio of the metal and carbon atom radii. For the latter, according to the Hägg's rule, if the metal-carbon radius ratio is from 0.41 to about 0.59 such interstitial close-packed phase might be formed by contrast with the ratio exceeding 0.59 where more complex structures typically appear. Therefore, among 3*d* transition metals, Sc, Ti, and V suppose the formation of interstitial rocksalt monocarbides while starting from Cr more complex compounds such as $Cr_{23}C_6$ [43] and $Fe_3C$ [7] are generally observed in the experiment. In addition, a recent theoretical study suggested the instability of *rs*-CoC, *rs*-NiC, *rs*-CuC both at zero temperature and any reasonable external conditions, and only possible formation of *rs*-FeC under pressure of 54 GPa. [30]

Abovementioned facts correlate with our further insight into found energetic superiority of *rs*-TMC (TM = Sc, Ti, V, Cr, Mn, and Fe) over the predicted tetragonal phase as shown in Figure 1 (blue region). Calculated transition pathways (see Figure 2a) from isolated *t*-TMC to the bulk phase, while monotonously reducing interlayer distance, have shown an absence of an energy barrier. Here, the direct transformation of *t*-TMC to either rocksalt phase (ScC, TiC) or *h*-TMC state (VC, CrC, MnC, FeC) was demonstrated. The latter, *h*-TMC, (see Figure S3) is similar to an hexagonal intermediate structure of AlN or GaN [44]. In particular, this phase was found to be metastable in the case of VC and further transformed to the energetically favorable rocksalt phase through a small energy barrier ~ 0.1 eV/VC, see Figure S4. By contrast, for CrC, MnC, and FeC, hexagonal phase has a lower energy than the corresponding rocksalt structure, however, the evidences of its thermal and kinetic stability are experimentally unknown to the best of our knowledge. Thus, in a full accordance with the mentioned Hägg's rule, some complex bulk structures of chromium, manganese and iron carbides are likely to be observed rather than metal monocarbides.

Finally, one can conclude that the studied *t*-TMC phase (TM = Sc, Ti, V, Cr, Mn, Fe) is stable only as isolated structure and, therefore, its formation might be possible under very specific conditions that intentionally confine its thickness. Moreover, even the bulk ground-state structure of mentioned monocarbides seems to be not entirely understood. Thus, further investigation of 3*d* bulk TMC formation under different conditions is of fundamental interest and might become the subject of future studies.

By contrast to two-dimensional *t*-TMC phase, the orthorhombic structures (*o*-CoC, *o*-NiC, *o*-CuC) as well as probably the border line case *o*-FeC are energetically more favorable than corresponding rocksalt bulk phase, see Figure 1 (green region). Some factors of the promising *o*-TMC stability have been recently discussed through the electronic structure and bonds pattern analysis. [22]

For these four 3*d* monocarbides we have designed a bulk unit cell consisting of AA-stacked weakly bounded *o*-TMC monolayers and calculated its energy as a function of the interlayer distance, see Figure 2b. For the calculated equilibrium interlayer distance of each structure the proposed AA-stacked bulk TMC has lower energy per TMC unit than the rocksalt structure, see Figure S5. In addition, we have compared 2D monolayers with their bulk structures and ultrathin films based on them. The binding energy, $E_{bind}$, was calculated as following:

$$E_{bind} = \frac{1}{N_{layer}}(E_{film} - N_{layer} \cdot E_{2D}), \qquad (1)$$

where $N_{layer}$ is the number of 2D monolayers in a film; $E_{film}$ and $E_{2D}$ are the energies of the corresponding *N*-layered film and one pristine 2D monolayer.

The calculated binding energy dependence as function of $1/N_{layer}$ for the four TMC is presented in Figure 2c.

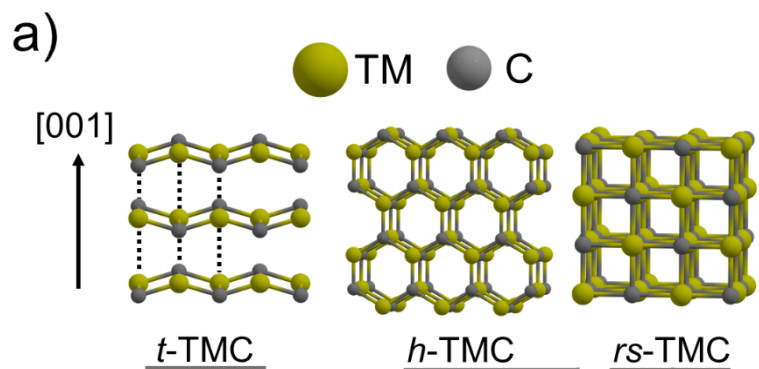
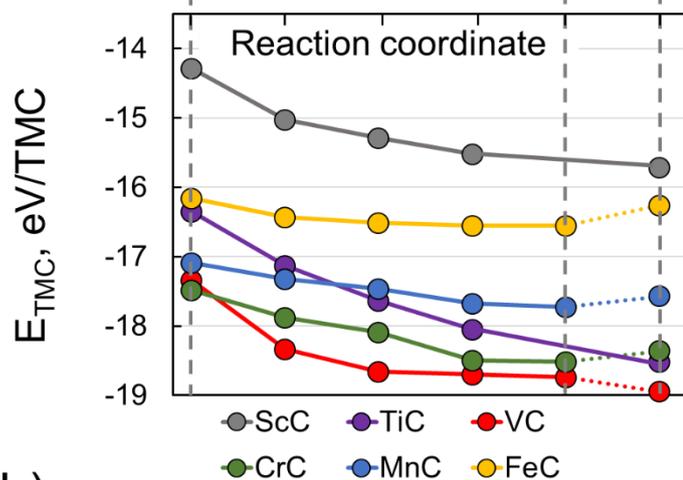
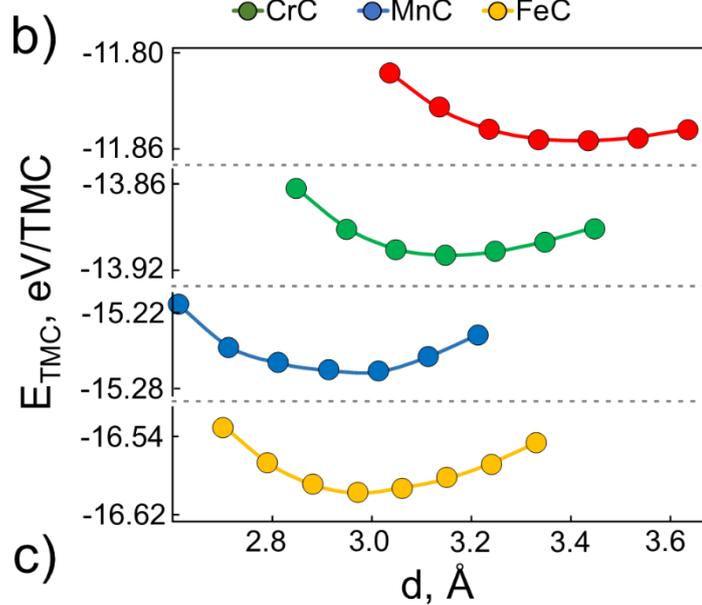
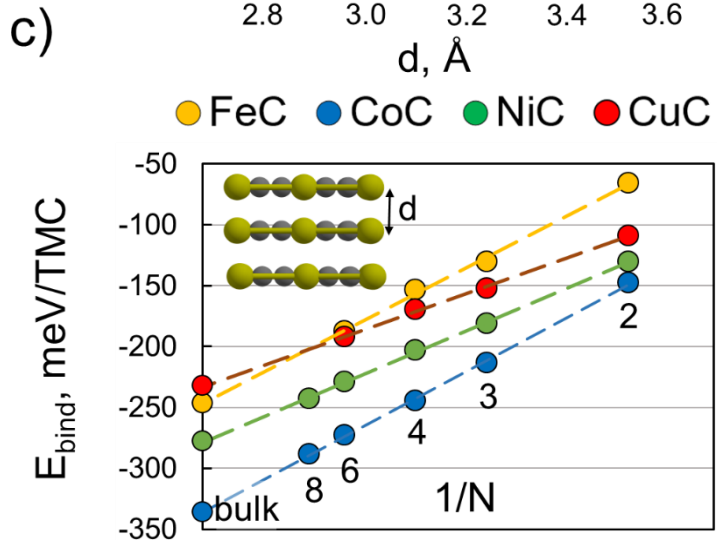

Figure 2. a) Transition pathways from isolated *t*-TMC to either *h*-TMC or *rs*-TMC bulk phase. Solid and dotted lines indicate absence or presence of some energy barrier, respectively. Side view on atomic configuration of *t*-TMC, *h*-TMC, and *rs*-TMC is shown above. b) Energy-distance dependences of AA-bulk of *o*-TMC monolayers. c) Dependence of interlayer binding energy in *o*-TMC film on $1/N_{layer}$. The structure model used in (b,c) is shown in the inset of figure (c).

One can see that opposite to the rocksalt bulk structure (see Figure 1 and Figure 2a), the orthorhombic bulk phase is energetically more favorable than isolated 2D monolayers. However, the difference is not large and is in the order of van der Waals binding energies (< 0.4 eV/TMC). Moreover, the binding energy between the layers of hypothetical ultrathin films consisting of only few orthorhombic layers is much smaller and reaches values of 70-150 meV/TMC. Therefore, if an orthorhombic bulk structure is synthesized, it might be cleaved into 2D layers.

Finally, one can conclude that the calculated energy difference between 2D *o*-TMC and 3D rocksalt phases (Figure 1) is likely irrelevant, because despite some theoretical predictions, the *rs*-phase have never been experimentally observed for these four monocarbides as discussed previously. Therefore, one can expect FeC, CoC, NiC, and CuC to either exist in some experimentally unknown bulk phase (i.e. other than rocksalt structure) or tend to form metal carbide ultrathin films and monolayers, e.g. *o*-TMC.

For further investigation of the layered structures stability, we studied variable compositions of 2D $TM_XC_Y$ compounds (TM = Fe, Co, Ni, Cu) via evolutionary algorithm as implemented in USPEX code. A main principle of this method is the consideration of all the possible decomposition lines and further drawing of its convex hull that indicates potentially stable structures. The formation energy of any $TM_XC_Y$ compound ($E_{form}$) can be calculated as:

$$E_{form} = \frac{1}{N_{TM} + N_C} [E_{TMC} - (N_{TM} \cdot E_{TM} + N_C \cdot E_C)], \qquad (2)$$

where $E_{TMC}$, $E_{TM}$, and $E_C$ are energies of transition metal carbide per unit cell, pure metal and pure carbon structures per atom, respectively; $N_{TM}$, $N_C$ are the number of transition metal and carbon atoms inside TMC unit cell, respectively.

However, a question about correctness of comparison of the structures with different dimensionality (e.g. 3D rocksalt phase and 2D monolayer) should be solved preliminarily. Indeed, since Eq. (2) includes energy of TMC monolayer together with energies of pristine metal and carbon films, all of them have to be calculated in the same "dimensional conditions".

For instance, multilayered stable two-dimensional structures allow a formation of new 2D films between layers by vertical confinement of atoms. Such "nanoreactor" was previously successfully used for the synthesis of two-dimensional ice between graphene/MoS$_2$ layers [45] as well as for the synthesis of two-dimensional lithium [10] and copper [46]. In a similar way, one may suggest 2D TMC to be thermodynamically stable with respect to disproportionation to the 2D adjacent phases: 2D carbon (graphene) and corresponding 2D metal.

Nevertheless, some variation of the 2D layer thickness is possible and should be taken into account. For instance, if we need to investigate a variable-composition convex hull of *only planar* monolayers such as *o*-TMC, we should also use energy of the most favorable planar carbon (that will be obviously graphene) and metal monolayers. However, if we wish to use less strict constrains and, therefore, take into account possible corrugation of a TMC film (e.g. *t*-TMC), we should increase the thickness limit of pristine carbon and metal film as well, while searching for the most favorable structures to be used in Eq. (2). Thus, usage of two different thickness thresholds (see Methods section) let us to consider only planar and possibly corrugated lattices.

Having based on the abovementioned, we have filtered all the found 2D TM$_X$C$_Y$ (TM = Fe, Co, Ni, Cu) and considered their formation energies. Figure 3a,b represent results of variable-composition calculations of TM$_X$C$_Y$ for planar and corrugates states, respectively, where dashed lines indicate convex hull of prospectively stable compounds. Any appearance of *o*-TMC (TM:C = 1:1), *t*-TMC (TM:C = 1:1) or planar tetracoordinated TMC (*ptC*-TMC, TM:C = 2:1, discussed below) is indicated by stars, squares and diamonds, respectively, and their atomic configurations are shown in Figure 3c.

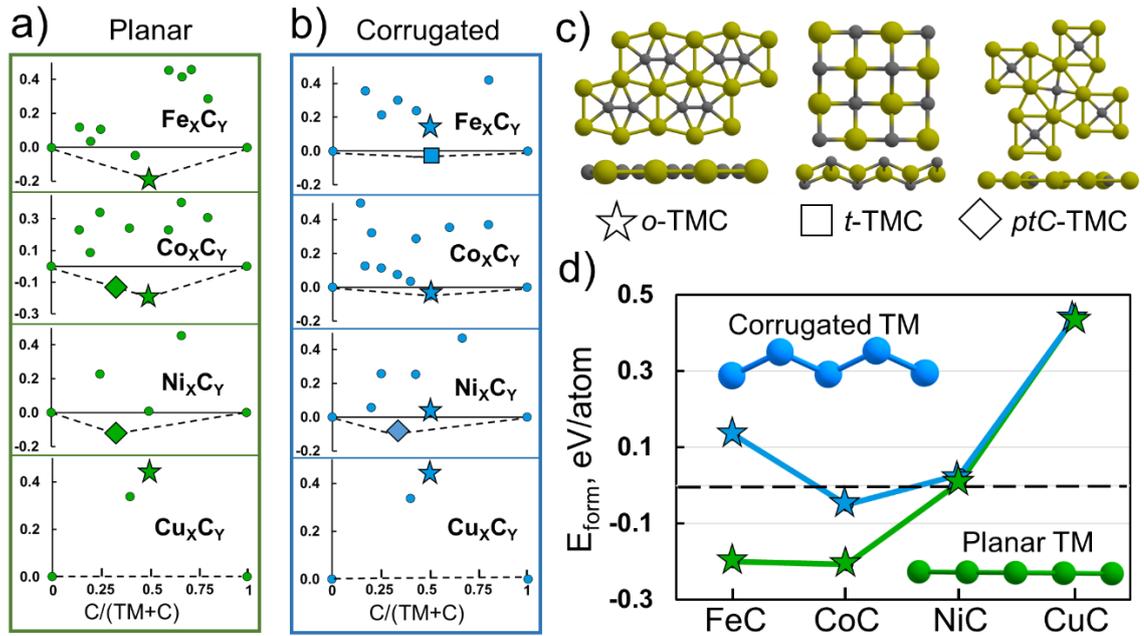

Figure 3. a,b) Variable-composition convex hulls of $TM_XC_Y$ (eV/atom) for only planar and possibly corrugated structures, respectively and c) atomic configurations of prospectively stable compounds: $o$-TMC (TM:C = 1:1), $t$-TMC (TM:C = 1:1), and $ptC$-TMC (TM:C = 2:1). d) Formation energy of $o$-TMC from (a,b). Atomic configurations of planar and corrugated metal films are shown in (d) by green and blue atoms, respectively. Everywhere, $o$-TMC, $t$-TMC, and $ptC$-TMC are indicated by stars, squares and diamonds, respectively.

First of all, from Figure 3a one can see that $o$-TMC is the most energetically favorable planar monolayer among other generated monocarbides, i.e. with TM:C = 1:1. Moreover, $o$-FeC and $o$-CoC are located at the convex hull (marked by dashed line which is lower than any possible decomposition line by its definition). This means that once vertical confinement is applied to FeC and CoC, they are expected to form planar monolayers rather than a metal/graphene interface. In contrast, the $o$-phase of NiC and CuC are metastable (see Figure 3a) and likely decompose to pure metal and carbon films (e.g. Cu/C interface). Nevertheless, in the case of $Ni_XC_Y$ another 2D phase with TM:C = 2:1 can appear (marked by diamond). This phase corresponds to so-called "planar tetracoordinated carbon" carbide ($ptC$-TMC) as proposed by Hoffmann et al. [47] and further studied in Ref. [48] by computational methods for $Co_2C$ and $Ni_2C$. In the case of $Cu_XC_Y$, no 2D phases are expected.

Afterwards, we have considered possible corrugation of 2D structures (Figure 3b). For $Fe_XC_Y$, the $t$-FeC phase appears to be stable in contrast to the metastable $o$-FeC. For $Co_XC_Y$, the stability of $o$-CoC is preserved at the convex hull, but now is closer to the zero decomposition line. The reason of such modifications is change of the energetically most favorable Fe and Co metal film from a planar monolayered in Figure 3a (see structure sketch in the lower right part of

Figure 3d) to a corrugated structure in Figure 3b (see structure sketch in the upper left part of Figure 3d). Indeed, while in the Eq. (2) $E_C$ refers to the energy of graphene (despite thickness limits), the change of $E_{TM}$ from the planar monolayered to the corrugated case will obviously change formation energies of every intermediate $TM_XC_Y$ (TM = Fe, Co) compound. In contrast, the formation energies of $Ni_XC_Y$ and $Cu_XC_Y$ are almost identical (see Figure 3a and Figure 3b). Indeed, the energetically favorable Ni monolayered film was found to be only slightly corrugated which energy is just a bit lower than for the perfect planar case. As for Cu, no corrugated copper monolayer was found and, therefore, in Figure 3a and Figure 3b the same $Cu_XC_Y$ points are presented.

Formation energies of *o*-TMC (indicated by stars in Figure 3a,b) are additionally shown in Figure 3d for better comparison. From there, one can see that *o*-FeC is expected to be stable only in comparison to the planar metal film and graphene (green color). Less strict vertical constraints (blue color) lead to the appearance of the *t*-FeC phase at the convex hull, which has already been considered as a metastable phase (see Figure 2a), unless being isolated as a free-standing monolayer. For *o*-CoC, the formation energy is negative, comparing with both planar and corrugated metal monolayer. In other words, *o*-CoC is supposed to be the most favorable 2D phase among any other 2D $Co_XC_Y$ structures. This corresponds well with the observation of Nevalaita et al.,[49] where pure metallic patches inside graphene pore and metal carbides were theoretically studied. Finally, in the case of NiC and CuC, their formation energies are positive and, therefore, these monocarbides are not expected to be formed as a monolayered film.

**CONCLUSIONS**

In summary, a systematic theoretical study on the formation of 3*d* transition metal monocarbides (TMC) from ScC to CuC was performed. By comparison of the energies of the conventional rocksalt bulk phase with possible 2D TMC phases, we have clearly shown that the energetically favorable 3D rocksalt phase is monotonously interchanged by recently reported planar orthorhombic phase characterized by carbon dimers inside metal hexagons. The lack of experimental data about synthesis of FeC, CoC, NiC, and CuC, in rocksalt phase correlates with our findings, that these monocarbides tend to exist either in another bulk state (e.g. we have proposed energy favorable AA stack of 2D orthorhombic layers) or appear as thermodynamically stable monolayers.

It was also predicted that 2D ground-state of the first part of 3*d* TMC (ScC, TiC, VC, CrC, MnC) is a corrugated tetragonal lattice, *t*-TMC. Nevertheless, such corrugated monolayers are intrinsically unstable as follows from calculated transition pathways from isolated monolayer to bulk phase, where the absence of an energy barrier is shown. Moreover, we have found that CrC, MnC, and FeC may be transformed to metastable hexagonal bulk phase.

Finally, by applying an evolutionary algorithm we have studied the formation possibility of a wide range of 2D $TM_XC_Y$ (TM = Fe, Co, Ni, Cu) compounds under the assumption of vertical confinement. It was shown that only *t*-FeC, *o*-FeC, *o*-CoC, *ptC*-$Co_2C$, and *ptC*-$Ni_2C$ are expected to be thermodynamically stable against other two-dimensional carbides and disproportionation into 2D carbon (graphene) and the corresponding 2D metal.

**CONFLICTS OF INTEREST**

There are no conflicts of interest to declare.

**ACKNOWLEDGEMENTS**

We thank Dr. Z.I. Popov for the comments that greatly improved the manuscript. The authors gratefully acknowledge the financial support of Russian Science Foundation (Project identifier: 17-72-20223). The calculations were performed at supercomputer cluster provided by the Materials Modeling and Development Laboratory at NUST "MISIS" and Joint Supercomputer Center of the Russian Academy of Sciences

**REFERENCES**

1  N. R. Glavin, R. Rao, V. Varshney, E. Bianco, A. Apte, A. Roy, E. Ringe and P. M. Ajayan, *Adv. Mater.*, 2019, 1904302.
2  N. Mounet, M. Gibertini, P. Schwaller, D. Campi, A. Merkys, A. Marrazzo, T. Sohier, I. E. Castelli, A. Cepellotti, G. Pizzi and N. Marzari, *Nat. Nanotechnol.*, 2018, **13**, 246–252.
3  J. Zhou, L. Shen, M. D. Costa, K. A. Persson, S. P. Ong, P. Huck, Y. Lu, X. Ma, Y. Chen, H. Tang and Y. P. Feng, *Sci. Data*, , DOI:10.1038/s41597-019-0097-3.
4  J. Yun, *Adv. Funct. Mater.*, 2017, **27**, 1606641.
5  S. T. Hunt, M. Milina, A. C. Alba-Rubio, C. H. Hendon, J. A. Dumesic and Y. Roman-Leshkov, *Science*, 2016, **352**, 974–978.
6  H. H. Hwu and J. G. Chen, *Chem. Rev.*, 2005, **105**, 185–212.
7  Y. Hu, J. O. Jensen, W. Zhang, L. N. Cleemann, W. Xing, N. J. Bjerrum and Q. Li, *Angew. Chem. Int. Ed.*, 2014, **53**, 3675–3679.
8  X. Fan, Z. Peng, R. Ye, H. Zhou and X. Guo, *ACS Nano*, 2015, **9**, 7407–7418.
9  W. Yang, X. Liu, X. Yue, J. Jia and S. Guo, *J. Am. Chem. Soc.*, 2015, **137**, 1436–1439.
10 M. Kühne, F. Börrnert, S. Fecher, M. Ghorbani-Asl, J. Biskupek, D. Samuelis, A. V. Krasheninnikov, U. Kaiser and J. H. Smet, *Nature*, 2018, **564**, 234–239.
11 X. Wang, C. Wang, C. Chen, H. Duan and K. Du, *Nano Lett.*, 2019, **19**, 4560–4566.
12 X. Zhao, J. Dan, J. Chen, Z. Ding, W. Zhou, K. P. Loh and S. J. Pennycook, *Adv. Mater.*, 2018, **30**, 1707281.
13 T. Joseph, M. Ghorbani-Asl, A. G. Kvashnin, K. V. Larionov, Z. I. Popov, P. B. Sorokin and A. V. Krasheninnikov, *J. Phys. Chem. Lett.*, 2019, **10**, 6492–6498.
14 J. Zhao, Q. Deng, A. Bachmatiuk, G. Sandeep, A. Popov, J. Eckert and M. H. Rümmeli, *Science*, 2014, **343**, 1228–1232.
15 Y. Shao, R. Pang and X. Shi, *J. Phys. Chem. C*, 2015, **119**, 22954–22960.

16 K. V. Larionov, D. G. Kvashnin and P. B. Sorokin, *J. Phys. Chem. C*, 2018, **122**, 17389–17394.
17 E. Kano, Dmitry. G. Kvashnin, Seiji. Sakai, Leonid. A. Chernozatonskii, Pavel. B. Sorokin, A. Hashimoto and M. Takeguchi, *Nanoscale*, 2017, **9**, 3980–3985.
18 D. Fan, S. Lu and X. Hu, *ArXiv180203673 Cond-Mat*.
19 Z. Zhang, X. Liu, B. I. Yakobson and W. Guo, *J. Am. Chem. Soc.*, 2012, **134**, 19326–19329.
20 B. Xu, H. Xiang, J. Yin, Y. Xia and Z. Liu, *Nanoscale*, 2018, **10**, 215–221.
21 K. V. Larionov, Z. I. Popov, M. A. Vysotin, D. G. Kvashnin and P. B. Sorokin, *JETP Lett.*, 2018, **108**, 13–17.
22 C. Zhu, H. Lv, X. Qu, M. Zhang, J. Wang, S. Wen, Q. Li, Y. Geng, Z. Su, X. Wu, Y. Li and Y. Ma, *J. Mater. Chem. C*, 2019, **7**, 6406–6413.
23 H. Nowotny and H. Auer-Welsbach, *Monatshefte Für Chem.*, 1961, **92**, 789–793.
24 K. Nakamura and M. Yashima, *Mater. Sci. Eng. B*, 2008, **148**, 69–72.
25 J. Wang, X. Chen, N. Yang and Z. Fang, *Appl. Phys. Solids Surf.*, 1993, **56**, 307–309.
26 B. X. Liu and X. Y. Cheng, *J. Phys. Condens. Matter*, 1992, **4**, L265–L268.
27 A. N. Arpita Aparajita, N. R. Sanjay Kumar, S. Chandra, S. Amirthapandian, N. V. C. Shekar and K. Sridhar, *Inorg. Chem.*, 2018, **57**, 14178–14185.
28 M. G. Quesne, A. Roldan, N. H. de Leeuw and C. R. A. Catlow, *Phys. Chem. Chem. Phys.*, 2018, **20**, 6905–6916.
29 J. S. Gibson, J. Uddin, T. R. Cundari, N. K. Bodiford and A. K. Wilson, *J. Phys. Condens. Matter*, 2010, **22**, 445503.
30 N. J. Szymanski, I. Khatri, J. G. Amar, D. Gall and S. V. Khare, *J. Mater. Chem. C*, 2019, **7**, 12619–12632.
31 P. E. Blöchl, *Phys. Rev. B*, 1994, **50**, 17953–17979.
32 G. Kresse and J. Furthmüller, *Comput. Mater. Sci.*, 1996, **6**, 15–50.
33 G. Kresse and J. Furthmüller, *Phys. Rev. B*, 1996, **54**, 11169–11186.
34 G. Kresse and J. Hafner, *Phys. Rev. B*, 1993, **47**, 558–561.
35 G. Kresse and J. Hafner, *Phys. Rev. B*, 1994, **49**, 14251–14269.
36 H. J. Monkhorst and J. D. Pack, *Phys. Rev. B*, 1976, **13**, 5188–5192.
37 S. Grimme, J. Antony, S. Ehrlich and H. Krieg, *J. Chem. Phys.*, 2010, **132**, 154104.
38 A. R. Oganov and C. W. Glass, *J. Chem. Phys.*, 2006, **124**, 244704.
39 A. O. Lyakhov, A. R. Oganov, H. T. Stokes and Q. Zhu, *Comput. Phys. Commun.*, 2013, **184**, 1172–1182.
40 F. Claeyssens, C. L. Freeman, N. L. Allan, Y. Sun, M. N. R. Ashfold and J. H. Harding, *J Mater Chem*, 2005, **15**, 139–148.
41 J. Häglund, A. Fernández Guillermet, G. Grimvall and M. Körling, *Phys. Rev. B*, 1993, **48**, 11685–11691.
42 U. Jansson and E. Lewin, *Thin Solid Films*, 2013, **536**, 1–24.
43 K. Hirota, K. Mitani, M. Yoshinaka and O. Yamaguchi, *Mater. Sci. Eng. A*, 2005, **399**, 154–160.
44 J. Cai and N. Chen, *Phys. Rev. B*, , DOI:10.1103/PhysRevB.75.134109.
45 P. Bampoulis, V. J. Teernstra, D. Lohse, H. J. W. Zandvliet and B. Poelsema, *J. Phys. Chem. C*, 2016, **120**, 27079–27084.
46 X.-C. Liu, S. Zhao, X. Sun, L. Deng, X. Zou, Y. Hu, Y.-X. Wang, C.-W. Chu, J. Li, J. Wu, F.-S. Ke and P. M. Ajayan, *Sci. Adv.*, 2020, **6**, eaay4092.
47 Roald. Hoffmann, R. W. Alder and C. F. Wilcox, *J. Am. Chem. Soc.*, 1970, **92**, 4992–4993.
48 E. Jimenez-Izal, M. Saeys and A. N. Alexandrova, *J. Phys. Chem. C*, 2016, **120**, 21685–21690.
49 J. Nevalaita and P. Koskinen, *Nanoscale*, 2019, **11**, 22019–22024.